\documentstyle[11pt,aasms4]{article}

\begin{document}

\title{The Visual Orbit of 64~Piscum}
\author{A.F.~Boden\altaffilmark{1,2},
	B.F.~Lane\altaffilmark{5},
	M.J.~Creech-Eakman\altaffilmark{5},
	M.M.~Colavita\altaffilmark{1},
	P.J. Dumont\altaffilmark{1},
	J.~Gubler\altaffilmark{3},
	C.D.~Koresko\altaffilmark{5},
	M.J.~Kuchner\altaffilmark{4},
	S.R.~Kulkarni\altaffilmark{4,5},
	D.W.~Mobley\altaffilmark{1},
	X.P.~Pan\altaffilmark{4},
	M.~Shao\altaffilmark{1},
	G.T.~van~Belle\altaffilmark{1},
	J.K.~Wallace\altaffilmark{1}\\
	(The PTI Collaboration),\\
	B.R.~Oppenheimer\altaffilmark{4}}
\altaffiltext{1}{Jet Propulsion Laboratory, California Institute of Technology}
\altaffiltext{2}{Infrared Processing and Analysis Center, California Institute of Technology}
\altaffiltext{3}{University of California, San Diego}
\altaffiltext{4}{Palomar Observatory, California Institute of Technology}
\altaffiltext{5}{Geology and Planetary Sciences, California Inststute of Technology}
\authoremail{bode@huey.jpl.nasa.gov}

\begin{abstract}
We report on the determination of the visual orbit of the double-lined
spectroscopic binary system 64 Piscum with data obtained by the
Palomar Testbed Interferometer in 1997 and 1998.  64~Psc is a nearly
equal-mass double-lined binary system whose spectroscopic orbit is
well known.  We have estimated the visual orbit of 64~Psc from our
interferometric visibility data.  Our 64~Psc orbit is in good
agreement with the spectroscopic results, and the physical parameters
implied by a combined fit to our interferometric visibility data and
radial velocity data of Duquennoy and Mayor result in precise
component masses that agree well with their spectral type
identifications.  In particular, the orbital parallax of the system is
determined to be 43.29 $\pm$ 0.46 mas, and masses of the two
components are determined to be 1.223 $\pm$ 0.021 M$_{\sun}$ and 1.170
$\pm$ 0.018 M$_{\sun}$, respectively.

Nadal et al.~put forward arguments of temporal variability in some of
the orbital elements of 64~Psc, presumably explained by an undetected
component in the system.  While our visibility data does not favor the
Nadal temporal variability inference, neither is it definitive in
excluding it.  Consequently we have performed both high dynamic-range
near-infrared imaging and spectroscopy of potential additional
companions to the 64 Psc system.  Our imaging and spectroscopic data
do not support the conjecture of an additional component to 64~Psc,
but we did identify a faint object with unusual red colors and
spectra.

\end{abstract}

\section{Introduction}

64 Piscum (HD 4676) is a nearby, short-period (13.8 d) binary system
with nearly-equal mass ($q$ $\sim$ 0.96) components of spectral type
F8 V.  Further, the 64~Psc binary is the bright component in an
apparent visual triple system (Washington Double Star Catalog -- WDS,
\cite{WDS97}).  The WDS lists two additional visual components B and C
with V of 12.6 and 13.0 at 77 and 71 arcsec separations from 64~Psc,
respectively.  If these objects are in fact bound to 64~Psc, with a
system distance of approximately 24 pc as determined by Hipparcos
(\cite{HIP97,Perryman97}), the minimum physical separations of the dim
B and C components to 64~Psc are approximately 1800 AU, indicating
orbital periods of $\sim$ 5 $\times$ 10$^4$ yr.

Abt and Levy (1976, herein AL76) determined the first double-lined
spectroscopic orbit for the 64~Psc system.  Both Nadal et al (1979,
herein N79) and later Duquennoy and Mayor (data in 1991a -- herein
DM91a, and solution in 1991b -- herein DM91b) have determined
additional double-lined orbital solutions for the 64~Psc system, which
are all in reasonable agreement with each other.  Particularly notable
is Nadal's argument for the potential presence of an additional,
undetected object in the 64~Psc system from possible time variation in
the system velocity, the time of periastron passage, and the argument
of periastron.  N79 used 1918 data from Lick and Mt.~Wilson,
reanalyzed the AL76 observations, and presented their own data in
support of the variation hypothesis.  Presumably these time-variation
effects could not be due to either the B or C visual companions listed
in the WDS; N79 estimates the period of these effects to be $\sim$
2700 yr.

Herein we report the determination of the 64~Psc visual orbit from
near-infrared, long-baseline interferometric measurements taken with
the Palomar Testbed Interferometer.  PTI is a 110-m $K$-band (2.2 $\mu$m)
interferometer located at Palomar Observatory, and described in detail
elsewhere (\cite{Colavita99a}).  PTI has a minimum fringe spacing of
roughly 4 mas at the sky position of 64~Psc, making this binary system
readily resolvable.

\section{Observations}
The interferometric observable used for these measurements is the
fringe contrast or {\em visibility} (squared) of an observed
brightness distribution on the sky.  Normalized in the interval [0:1],
a single star exhibits visibility modulus in a uniform disk model
given by:
\begin{equation}
V =
\frac{2 \; J_{1}(\pi B \theta / \lambda)}{\pi B \theta / \lambda}
\label{eq:V_single}
\end{equation}
where $J_{1}$ is the first-order Bessel function, $B$ is the projected
baseline vector magnitude at the star position, $\theta$ is the
apparent angular diameter of the star, and $\lambda$ is the
center-band wavelength of the interferometric observation.  The
expected squared visibility in a narrow pass-band for a binary star
such as 64~Psc is given by:
\begin{equation}
V^{2} = \frac{V_{1}^2 + V_{2}^2 \; r^2
	        + 2 \; V_{1} \; V_{2} \; r \;
	          \cos(\frac{2 \pi}{\lambda} \; {\bf {B}} \cdot {\bf {s}})}
	      {(1 + r)^2}
\label{eq:V2_double}
\end{equation}
where $V_{1}$ and $V_{2}$ are the visibility moduli for the two stars
separately as given by Eq.~\ref{eq:V_single}, $r$ is the apparent
brightness ratio between the primary and companion, ${\bf {B}}$ is the
projected baseline vector at the system sky position, and ${\bf {s}}$
is the primary-secondary angular separation vector on the plane of the
sky (\cite{Pan90,Hummel95}).  The $V^2$ observables used in our 64~Psc
study are both narrow-band $V^2$ from individual spectral channels
(\cite{Colavita99a}), and a synthetic wide-band $V^2$, given by an
incoherent SNR-weighted average $V^2$ of the narrow-band channels in
the PTI spectrometer (\cite{Colavita99b}).  In this model the expected
wide-band $V^2$ observable is approximately given by an average of the
narrow-band formula over the finite pass-band of the spectrometer:
\begin{equation}
V^{2}_{wb} = \frac{1}{n}\sum_{i}^{n} V^{2}_{nb-i}(\lambda_i)
\label{eq:V2_doubleWB}
\end{equation}
where the sum runs over the channels covering the infrared K-band (2 -
2.4 $\mu$m) of the PTI spectrometer (\cite{Colavita99a}).  Separate
calibrations and hypothesis fits to the narrow-band and synthetic
wide-band $V^2$ datasets yield statistically consistent results, with
the synthetic wide-band data exhibiting superior fit performance.
Consequently we will present only the results from the synthetic
wide-band data.

64~Psc was observed in conjunction with our selected calibrator list
by PTI on 21 nights between 30 Aug 1997 and 16 Nov 1998, covering
roughly 32 periods of the system.  64~Psc, along with calibration
objects, was observed multiple times during each of these nights, and
each observation, or scan, was approximately 120 sec long.  For each
scan we computed a mean $V^2$ value from the scan data, and the error
in the $V^2$ estimate from the RMS internal scatter
(\cite{Colavita99b}).  64~Psc was always observed in combination with
one or more calibration sources within $\sim$ 20 degrees on the sky.
For our study we have used three late-type main sequence stars as
calibration objects: HD 166, HD 3651, and HD 4628.  Table
\ref{tab:calibrators} lists the relevant physical parameters for the
calibration objects.

The calibration of 64~Psc $V^2$ data is performed by estimating the
interferometer system visibility ($V^{2}_{sys}$) using calibration
sources with model angular diameters, and then normalizing the raw
64~Psc visibility by $V^{2}_{sys}$ to estimate the $V^2$ measured by
an ideal interferometer at that epoch (\cite{Mozurkewich91,Boden98}).
Calibrating our 64~Psc dataset with respect to the three calibration
objects listed in Table \ref{tab:calibrators} results in a total of 88
calibrated scans on 64~Psc over 21 nights in 1997 and 1998.

\begin{table}[t]
\begin{center}
\begin{small}
\begin{tabular}{|c|c|c|c|c|}
\hline
Object    & Spectral & Star        & 64~Psc        & Adopted Model \\
Name      & Type     & Magnitude   & Separation    & Diameter (mas)  \\
\hline
HD 166    & K0 V     & 6.1 V/4.1 K & 16$^{\circ}$  & 0.66 $\pm$ 0.06   \\
HD 3651   & K0 V     & 5.9 V/3.9 K & 4.9$^{\circ}$ & 0.79 $\pm$ 0.05   \\
HD 4628   & K2 V     & 5.7 V/3.5 K & 12$^{\circ}$  & 0.92 $\pm$ 0.05   \\
\hline
\end{tabular}
\caption{PTI 64~Psc Calibration Objects Considered in our Analysis.
The relevant parameters for our two calibration objects are
summarized.  The apparent diameter values are determined from
effective temperature and bolometric flux estimates based on archival
broad-band photometry, and visibility measurements with PTI.
\label{tab:calibrators}}
\end{small}
\end{center}
\end{table}

\section{Orbit Determination}

The estimation of the 64~Psc visual orbit is made by fitting a
Keplarian orbit model with visibilities predicted by
Eqs.~\ref{eq:V2_double} and \ref{eq:V2_doubleWB} directly to the
calibrated (narrow-band and synthetic wide-band) $V^2$ data on 64~Psc
(see \cite{Armstrong92b,Hummel93,Hummel95,Boden99}).  The fit is
non-linear in the Keplarian orbital elements, and is therefore
performed by non-linear least-squares methods (i.e.~the
Marquardt-Levenberg method, \cite{Press92}).  As such, this fitting
procedure takes an initial estimate of the orbital elements and other
parameters (e.g. component angular diameters, brightness ratio), and
evolves that model into a new parameter set which best fits the data.
However, the chi-squared surface has many local minima in addition to
the global minimum corresponding to the true orbit.  Because the
Marquardt-Levenberg method strictly follows a downhill path in the
$\chi^2$ manifold, it is necessary to thoroughly survey the space of
possible binary parameters to distinguish between local minima and the
true global minimum.  In addition, as the $V^2$ observable for the
binary (Eqs.~\ref{eq:V2_double} and \ref{eq:V2_doubleWB}) is invariant
under a rotation of 180$^{\circ}$, we cannot differentiate between an
apparent primary/secondary relative orientation and its mirror image
on the sky.  Consequently there remains a 180$^{\circ}$ ambiguity in
our determination of the longitude of the ascending node, $\Omega$,
which we quote by convention in the interval [0:180).  By similar
arguments our $V^2$ observable does not distinguish the longitude of
periastron ($\omega$) for the primary and secondary component.  We
have quoted our estimate constrained to be grossly (within
180$^{\circ}$) consistent with the consensus $\omega_1$ value of
approximately 200$^{\circ}$ (AL76, N79, DM91b).

In addition to our PTI visibility data we have used the double-lined
radial velocity data from DM91a.  To incorporate this data we have
upgraded our orbit estimation software to utilize both interferometric
visibility and radial velocity data either separately or
simultaneously (see similar remarks in \cite{Hummel98}).  The
$\omega$-degeneracy discussed above is resolved by the inclusion of
radial velocity data in our orbital solution (Table~\ref{tab:orbit}),
however the determination of $\Omega$ remains ambiguous by
180$^\circ$.

In the case of 64~Psc the parameter space is significantly narrowed by
the high-quality spectroscopic orbit from both N79 and DM91b, and the
Hipparcos distance determination sets the rough scale of the
semi-major axis (\cite{HIP97}).  Further, because the 64~Psc system is
nearly equal-mass ($q$ = 0.959 $\pm$ 0.004 -- DM91b), and because
interferometric $V^2$ measurements are only weakly sensitive to
variations in intensity ratio near equal intensity, we have
constrained the intensity ratios in our fits to 0.90 ($\Delta K$ =
0.11) as indicated by our component mass solutions (\S
\ref{sec:physics}), and the empirical mass-luminosity relationship of
Henry and McCarthy (1992).  Finally, at the distance of 64~Psc the
apparent diameters of the two main-sequence components of the system
are not strongly resolved by PTI, so we have constrained the estimated
diameters of both components, to 0.5 mas for the primary component and
0.45 mas for the secondary component, as indicated by our system
distance and component mass estimates (\S \ref{sec:physics}), and a
fit to empirical mass-radius data for main sequence eclipsing systems
given in Andersen (1991).  Given this limited parameter space, the
correct orbit solution is readily obtained by exhaustive search.

\begin{figure}
\epsscale{0.7}
\plotone{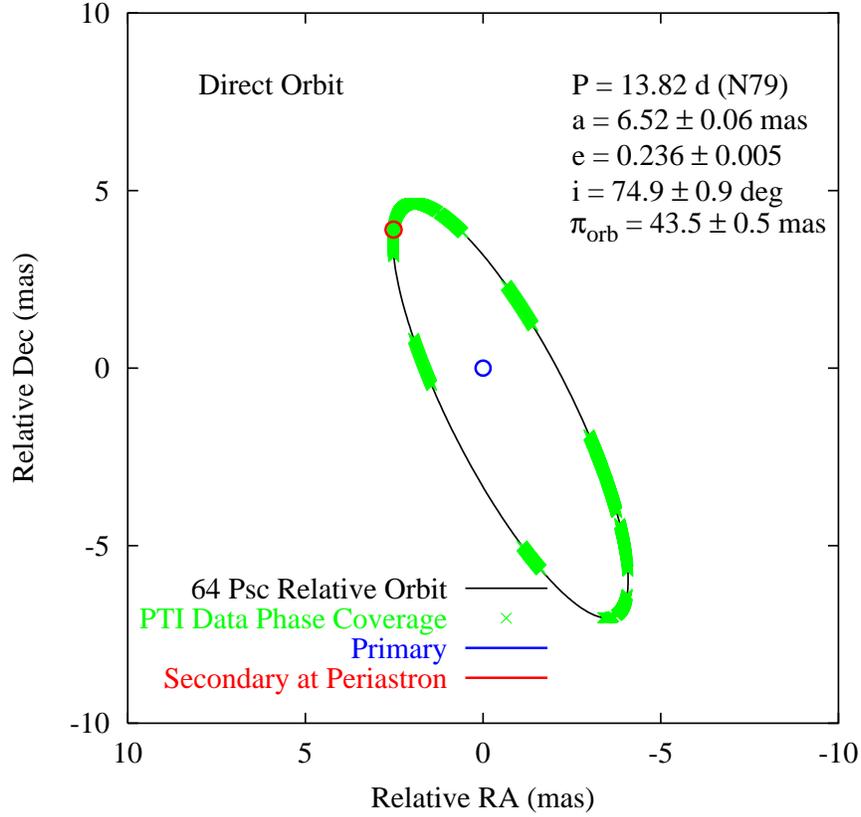}
\caption{Visual Orbit of 64~Psc.  The relative visual orbit of 64~Psc
is shown, with the primary and secondary objects rendered at T$_0$
(periastron).  The heavy lines along the relative orbit indicate areas
where we have orbital phase coverage in our PTI data (they are not
separation vector estimates); our data samples most phases of the
orbit well, leading to a reliable orbit determination.
\label{fig:64p_orbit}}
\end{figure}

Figure \ref{fig:64p_orbit} depicts the relative visual orbit of the
64~Psc system, with the primary component rendered at the origin, and
the secondary component rendered at periastron.  We have indicated the
phase coverage of our $V^2$ data on the relative orbit with heavy
lines; our data samples most phases of the orbit well, leading to a
reliable orbit determination.

\begin{figure}
\epsscale{0.8}
\plotone{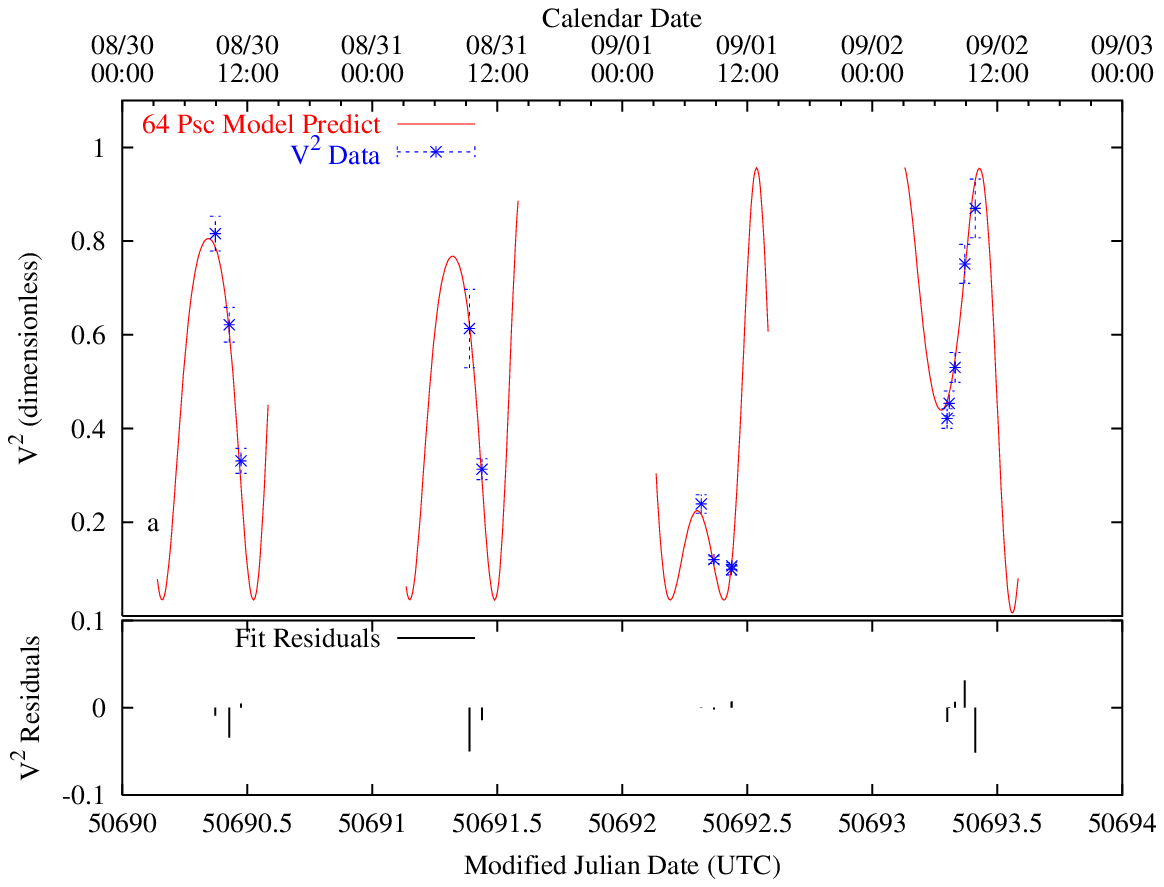}\\
\plotone{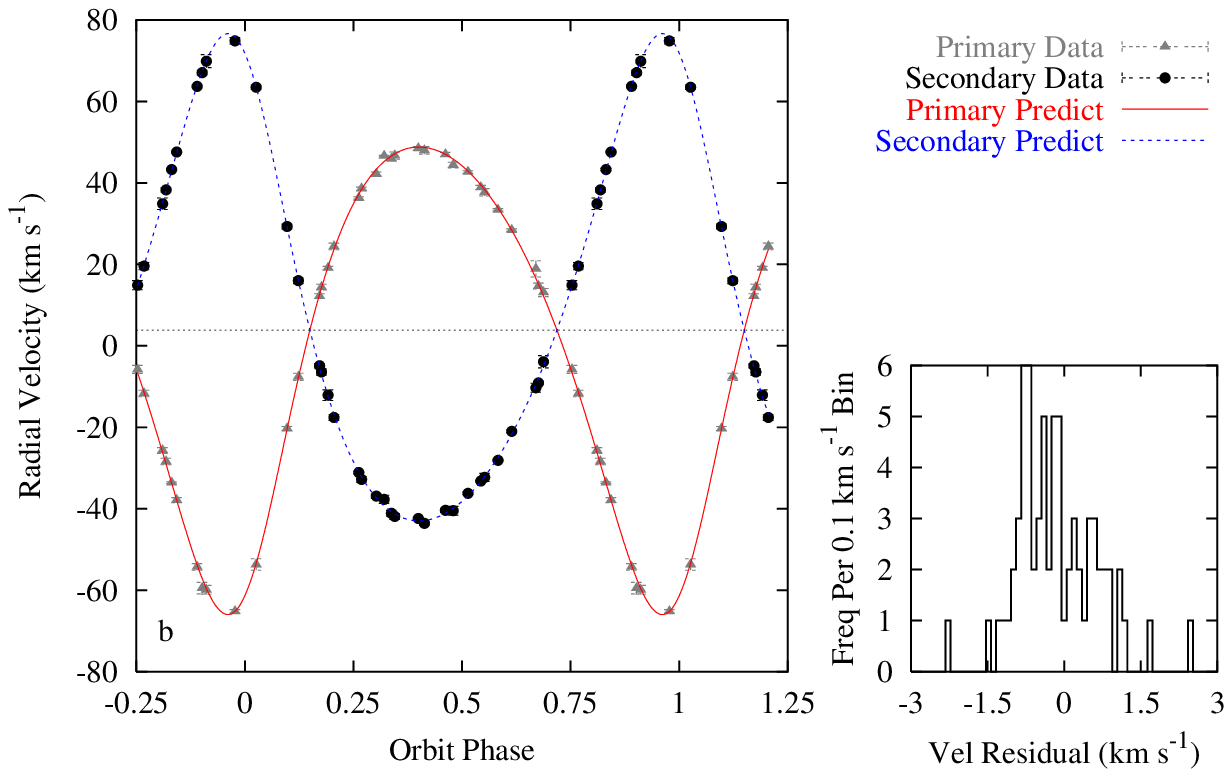}
\caption{$V^2$ and RV Fit of 64~Psc.  a: Four consecutive nights (30
Aug -- 2 Sept 1997) of calibrated $V^2$ data on 64~Psc, and $V^2$
predictions from the best-fit model for the system.  The lower frame
shows individual $V^2$ residuals between the calibrated data and
best-fit model.  b: A phase plot of radial velocity data from DM91a
and fit predictions from our Full-Fit solution
(Table~\ref{tab:orbit}).  Also included is a histogram of RV residuals
around the double-lined solution.
\label{fig:64p_fit}}
\end{figure}

Table \ref{tab:dataTable} lists the complete set of $V^2$ measurements
in our 64~Psc dataset and the prediction based on the best-fit orbit
model (our full-fit model, Table \ref{tab:orbit}) for 64~Psc.  Table
\ref{tab:dataTable} gives $V^2$ measurements and times, measurement
errors, model predictions, the photon-weighted average wavelength,
$u-v$ coordinates, and on-target hour angle for each of our calibrated
64~Psc observations.  Figure \ref{fig:64p_fit}a shows four consecutive
nights of PTI $V^2$ data on 64~Psc (30 Aug -- 2 Sept 1997), and $V^2$
predictions based on the best-fit model for the system (our full-fit
model, Table \ref{tab:orbit}).  The model predictions are in good
agreement with the observed data, with an RMS $V^2$ residual of 0.034
(average absolute $V^2$ deviation of 0.053), and a $\chi^2$ per Degree
of Freedom (DOF) of 0.85.  Figure \ref{fig:64p_fit}b gives a radial
velocity phase plot of the DM91a radial velocity data and the
predictions of our full-fit orbital solution.  A radial velocity fit
residual histogram is also included.

\begin{table}
\dummytable\label{tab:dataTable}
\end{table}

Spectroscopic orbit parameters (from N79 and DM91b) and our visual and
spectroscopic orbit parameters of the 64~Psc system are summarized in
Table \ref{tab:orbit}.  We give the results of separate fits to only
our $V^2$ data (our $V^2$-only Fit solution), and a simultaneous fit to
our $V^2$ data and the double-lined radial velocities from DM91a (our
Full-Fit solution) -- both with relative intensity and component
diameters constrained as noted above.  For the orbit parameters we
have estimated from our visibility data we list a total one-sigma
error in the parameter estimate, and the separate one-sigma errors in
the parameter estimates from statistical (measurement uncertainty) and
systematic error sources.  In our analysis the dominant forms of
systematic error are: (1) uncertainties in the calibrator angular
diameters (Table \ref{tab:calibrators}); (2) uncertainty in our
center-band operating wavelength ($\lambda_0 \approx$ 2.2 $\mu$m),
which we have taken to be 20 nm ($\sim$1\%); (3) the geometrical
uncertainty in our interferometric baseline ( $<$ 0.01\%); and (4)
uncertainties in orbital parameters we have constrained in our fitting
procedure (i.e.~the component intensity ratio and angular diameters in
both solutions).  Different parameters are affected differently by
these error sources; e.g.~our estimated uncertainty in the 64~Psc
orbital inclination is dominated by measurement uncertainty, while the
uncertainty in the angular semi-major axis is dominated by uncertainty
in the wavelength scale.  Conversely, we have assumed that all of the
error quoted by N79 and DM91b in the 64~Psc spectroscopic parameters
is statistical.

\begin{table}
\begin{center}
\begin{small}
\begin{tabular}{|c|c|c||c|c|}
\hline
Orbital			& N79     & DM91b    	& \multicolumn{2}{c|}{PTI 97/98} \\
\cline{4-5}
Parameter       	&         &         	& $V^2$-only Fit   & Full Fit \\
\hline \hline
Period (d)              & 13.824581 & 13.8318   & 13.82407      & 13.824621          \\
                        & $\pm$ 4.0 $\times$ 10$^{-5}$ & $\pm$ 1.7 $\times$ 10$^{-3}$ & $\pm$ 4.4 (3.8/2.2) $\times$ 10$^{-4}$ & $\pm$ 1.7 (1.7/0.4) $\times$ 10$^{-5}$     \\
T$_{0}$ (MJD)           & 41933.702 & 43468.308 & 50905.934 & 50905.984 \\
                        & $\pm$ 0.074 & $\pm$ 0.018  & $\pm$ 0.068 (0.048/0.048) &  $\pm$ 0.015 (0.015/0.003) \\
$e$                     & 0.243 $\pm$ 0.010 & 0.238 $\pm$ 0.002  &   0.2348    & 0.2376 \\
			&                   &                    & $\pm$ 0.0052 (0.0035/0.0039) & $\pm$ 0.0012 (0.0011/0.0005) \\
K$_1$ (km s$^{-1}$)     & 57.53 $\pm$ 0.74 & 57.31 $\pm$ 0.19          &  & 57.35 $\pm$ 0.31 (0.31/0.01) \\
K$_2$ (km s$^{-1}$)     & 58.77 $\pm$ 0.75 & 59.77 $\pm$ 0.18          &  & 59.95 $\pm$ 0.32 (0.31/0.02) \\
$\omega_{1}$ (deg)      & 199.6 $\pm$ 2.2  & 202.3 $\pm$ 0.5           &  202.2 $\pm$ 1.9 (1.3/1.3) & 203.56 $\pm$ 0.35 (0.33/0.10) \\
$\Omega_{1}$ (deg) &    & & 63.43 $\pm$ 0.68 (0.66/0.16) & 63.60 $\pm$ 0.82 (0.81/0.16) \\
$i$ (deg) &             & & 74.37 $\pm$ 0.96 (0.92/0.27) & 73.80 $\pm$ 0.92 (0.91/0.17) \\
$a$ (mas) &             & & 6.516 $\pm$ 0.061 (0.016/0.059) & 6.527 $\pm$ 0.061 (0.015/0.059) \\
$\Delta K$ (mag) & &    & {\em 0.11} & {\em 0.11} \\
$\chi^2$/DOF	      &  & & 0.85 & 1.2 (0.95 $V^2$/2.8 RV)  \\
$\overline{|R_{V^2}|}$ & &  & 0.053 & 0.053 \\
$\overline{|R_{RV}|}$ (km s$^{-1}$)  & & 0.65 &        & 0.65  \\
\hline
\end{tabular}
\end{small}
\caption{Orbital Parameters for 64~Psc.  Summarized here are the
apparent orbital parameters for the 64~Psc system as determined by
N79, DM91b, and PTI.  We give two separate fits to our data, with and
without including the double-lined DM91a radial velocities in the fit
(constrained parameters are listed in italics).  For parameters we
have estimated from including our PTI observations we separately quote
one-sigma errors from both statistical and systematic sources, given
as ($\sigma_{stat}$/$\sigma_{sys}$), and the total error as the sum of
the two in quadrature.  We have quoted the longitude of the ascending
node parameter ($\Omega$) as the angle between local East and the
orbital line of nodes measured positive in the direction of local
North.  Due to the degeneracy in our $V^2$ observable there is a
180$^\circ$ ambiguity in $\Omega$; by convention we quote it in the
interval of [0:180).  We quote mean absolute $V^2$ and RV residuals in
the fits, $\overline{|R_{V^2}|}$ and $\overline{|R_{RV}|}$
respectively.
\label{tab:orbit}}
\end{center}
\end{table}

There is a significant ($\sim$ 4.2-sigma) difference in the period as
estimated by DM91b and the period estimates of N79 and PTI.  For the
full-fit solution our period estimate exploits the roughly 7300-day
(530 64~Psc orbital cycles) time baseline between the DM91a RV data
and our own visibility data.  Both our period estimates agree well
with the N79 determination (the full-fit solution agreement with N79
is $\sim$ 0.9-sigma).

\section{Physical Parameters}
\label{sec:physics}
Physical parameters derived from our 64~Psc full-fit
visual/spectroscopic orbit are summarized in Table \ref{tab:physics}.
As in Table \ref{tab:orbit}, for physical parameters we have estimated
we quote total sigma error, and its statistical and systematic
contributions.  Notable among these is the high-precision
determination of the component masses for the system, a virtue of the
precision of the DM91a radial velocity measurements on both components
and the high inclination of the orbit.  We estimate the masses of the
F8V primary and secondary components as 1.223 $\pm$ 0.021 M$_{\sun}$
and 1.170 $\pm$ 0.018 M$_{\sun}$, respectively.  Remarkably, these are
very near the values estimated by N79, based on their radial
velocities and the relative magnitude in the system.

The Hipparcos catalog lists the parallax of 64~Psc as 41.80 $\pm$ 0.75
mas (\cite{HIP97}).  The distance determination to 64~Psc based on our
orbital solution is 23.10 $\pm$ 0.24 pc, corresponding to an orbital
parallax of 43.29 $\pm$ 0.46 mas, consistent with the Hipparcos result
at 3.5\% and 1.7-sigma.

Based on eclipsing binary systems, Popper (1980) and Andersen (1991)
list main-sequence linear diameters for stars in the mass range of the
64~Psc components at 1.15 -- 1.25 R$_{\sun}$.  At a distance of
approximately 23 pc this corresponds to apparent angular diameters in
the range of 0.42 -- 0.5 mas.  This range of apparent diameters is
unresolved to our interferometer, with expected component visibility
moduli (Eq.~\ref{eq:V_single}) of approximately 0.99.  Consequently,
as discussed above we have adopted model diameters for the two
components based on a fit to Andersen's compilation of eclipsing
binary data in this mass range.  In particular we have constrained our
solutions to a primary diameter of 0.5 mas (1.24 R$_{\sun}$ @ 23 pc)
and a secondary diameter 0.45 (1.12 R$_{\sun}$ @ 23 pc) respectively,
and included 10\% one-sigma errors around these constraints into our
systematic error computations.  We have also ignored any corrections
due to stellar limb darkening; our data would be highly insensitive to
these effects.

Similarly, because the system is nearly equal mass, hence nearly equal
brightness, our $V^2$ data is relatively insensitive to the magnitude
difference of the two components (see \cite{Hummel98} for similar
comments).  Based on our masses and the mass-luminosity models of
Henry \& McCarthy (1992, 1993) one would expect a $K$-magnitude
difference of 0.11 and a $V$-magnitude difference of 0.24.  For
reference, N79 quoted a best-fit component magnitude difference in the
visible of 0.16 without a quoted error.  An independent fit of our PTI
data for a K-band magnitude difference between the two yields
components 0.141 $\pm$ 0.076 -- well within one sigma of the expected
$K$-band magnitude difference.  As noted above for our orbit fit we
have chosen to constrain the $K$-magnitude difference at its model
value, and included a 5\% one-sigma error around this constraint in
our systematic error computations.  However, our fit K-band magnitude
difference, system distance estimate, and a 64~Psc apparent (CIT)
K-band magnitude estimate of 3.88 $\pm$ 0.07 yields K-band absolute
magnitude estimates of 2.75 $\pm$ 0.08 and 2.89 $\pm$ 0.09 for the
primary and secondary components respectively.  These estimates agree
well with the predictions of the mass-luminosity model of Henry and
McCarthy (\cite{Henry92}), which predicts absolute K-band magnitudes
of 2.79 and 2.91 for the primary and secondary components
respectively.

\begin{table}
\begin{center}
\begin{small}
\begin{tabular}{|c|c|c|}
\hline
Physical	 & Primary           & Secondary \\
Parameter        & Component         & Component \\
\hline \hline
a (10$^{-2}$ AU) & 7.371 $\pm$ 0.052 (0.052/0.006)   & 7.706 $\pm$ 0.054 (0.054/0.07)  \\
Mass (M$_{\sun}$)& 1.223 $\pm$ 0.021 (0.021/0.002)   & 1.170 $\pm$ 0.018 (0.018/0.002)  \\
Sp Type (DM91b)   & F8V              & F8V            \\
Model Diameter (mas) & 0.5 (const)   & 0.45 (const)     \\
\cline{2-3}
System Distance (pc) & \multicolumn{2}{c|}{23.10 $\pm$ 0.24 (0.13/0.21)} \\
$\pi_{orb}$ (mas)    & \multicolumn{2}{c|}{43.29 $\pm$ 0.46 (0.24/0.39)} \\
\cline{2-3}
M$_K$ (mag)      & 2.75 $\pm$ 0.08 (0.07/0.04)   & 2.89 $\pm$ 0.09 (0.08/0.04)  \\
\hline
\end{tabular}
\end{small}
\caption{Physical Parameters for 64~Psc.  Summarized here are the
physical parameters for the 64~Psc system as derived from the orbital
parameters in Table \ref{tab:orbit}.  As for our PTI-derived orbital
parameters we have quoted both total error and separate contributions
from statistical and systematic sources (given as
$\sigma_{stat}$/$\sigma_{sys}$).
\label{tab:physics}}
\end{center}
\end{table}

\section{Temporal Variations in Orbital Elements}

N79 argued for the possible presence of an additional, undetected
object in the 64~Psc system from supposed temporal variations in the
systematic velocity, the time of periastron (T$_0$), and the argument
of (primary) periastron ($\omega_1$).  In particular, N79 gives plots
of systematic velocity, projected T$_0$, and $\omega_1$ as a function
of observation epoch.  Assuming the N79 estimate for the period of the
effects (~2700 yr), the approximate angular separation between 64~Psc
and this third component would be $\sim$ 10 arcsec, sufficiently
distant to not affect our $V^2$ measurements on the inner pair, and
thus not perturb our orbit reconstruction.

\begin{figure}
\epsscale{0.7}
\plotone{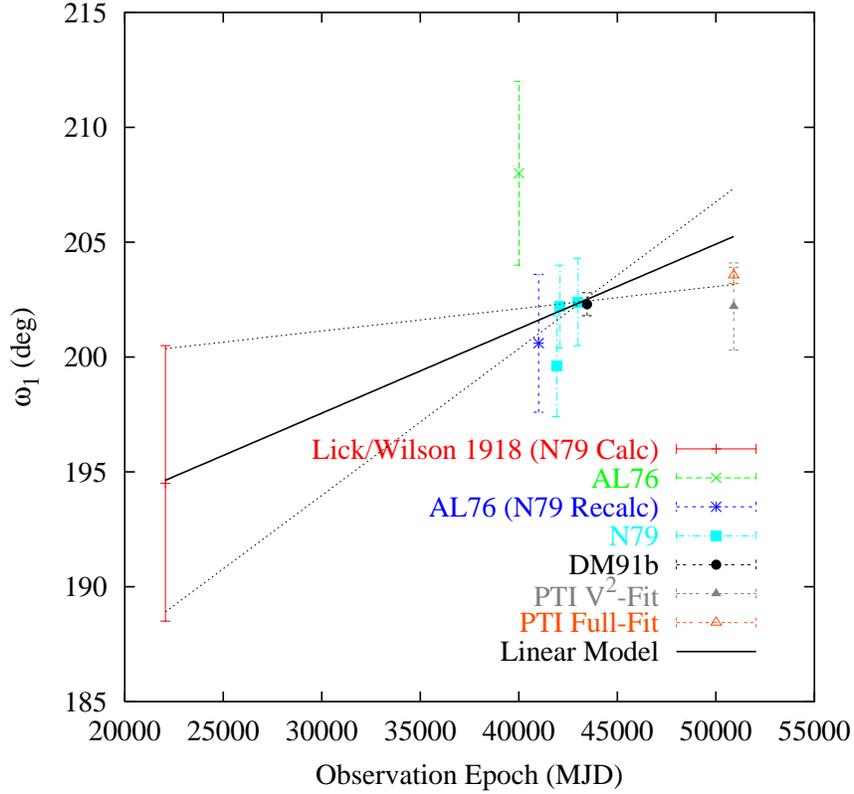}
\caption{Measurements of $\omega_1$ vs.~Time for 64~Psc.  At least as
calculated by N79, measurements of $\omega_1$ vs.~epoch of observation
show a striking linear trend, and N79 suggest the presence of an
additional, undetected body in the 64~Psc system to explain these
effects.  The N79 linear model for $\omega_1$ (updated here to include
the DM91b determination, but not the PTI result) is not in
particularly good agreement with the value for $\omega_1$ estimated
from our 1997/1998 PTI observations of 64~Psc (both separately and in
combination with the DM91a radial velocity measurements).  Our
combined interferometry-radial velocity determination of $\omega_1$ is
sufficiently accurate to formally exclude the N79 evolution hypothesis
at high confidence, but it conceivably may be biased by the older
radial velocities.
\label{fig:omega}}
\end{figure}

Our interferometric data does not address the 64~Psc systematic
velocity, but we can test the consistency of the N79 orbital element
trends with our orbit solution.  Because we note a bias in the period
estimate could produce the linear effects in projected T$_0$ that N79
claims, the comparison is best done with $\omega_1$.  Figure
\ref{fig:omega} gives a plot of various values of $\omega_1$ as
reported by AL76, N79, DM91b, and PTI.  Updating the N79 model for
$\omega_1$, we have fit a linear model to $\omega_1$ based on 1918
measurements made at Lick and Mt. Wilson (reduced by N79), AL76
(reanalyzed by N79), N79 themselves, and DM91b.  As depicted in Figure
\ref{fig:omega}, this model is dominated by the estimate determined by
N79 from the 1918 Lick/Wilson data, and based on the errors quoted by
N79 the data will support a broad range of trends for the
time-variation of $\omega_1$ -- including essentially no variation.
The updated N79 model for $\omega_1$ evaluated at the epoch of our
1997/1998 PTI observations is not in particularly good agreement with
our $V^2$-fit and full-fit values for $\omega_1$ of 202.7 $\pm$ 1.9
and 203.56 $\pm$ 0.35 respectively (Table \ref{tab:orbit}); it
predicts a value for $\omega_1$ of 205.0 $\pm$ 2.0 deg.  Shown in
Figure~\ref{fig:omega}, the disagreement (4-sigma) of our full-fit
$\omega_1$ determination to this linear model central projection would
seem to cast doubt on the evolving-$\omega_1$ hypothesis at a
reasonably high confidence level.  The potential problem with drawing
a firm conclusion on the evolution question is the inclusion of the
older DM91a radial velocities -- while the RV data match our recent
visibility measurements very well, they may well bias the $\omega_1$
estimate away from its current true value.

\section{Companion Search}
\begin{figure}[p]
\epsscale{1.0}
\plotone{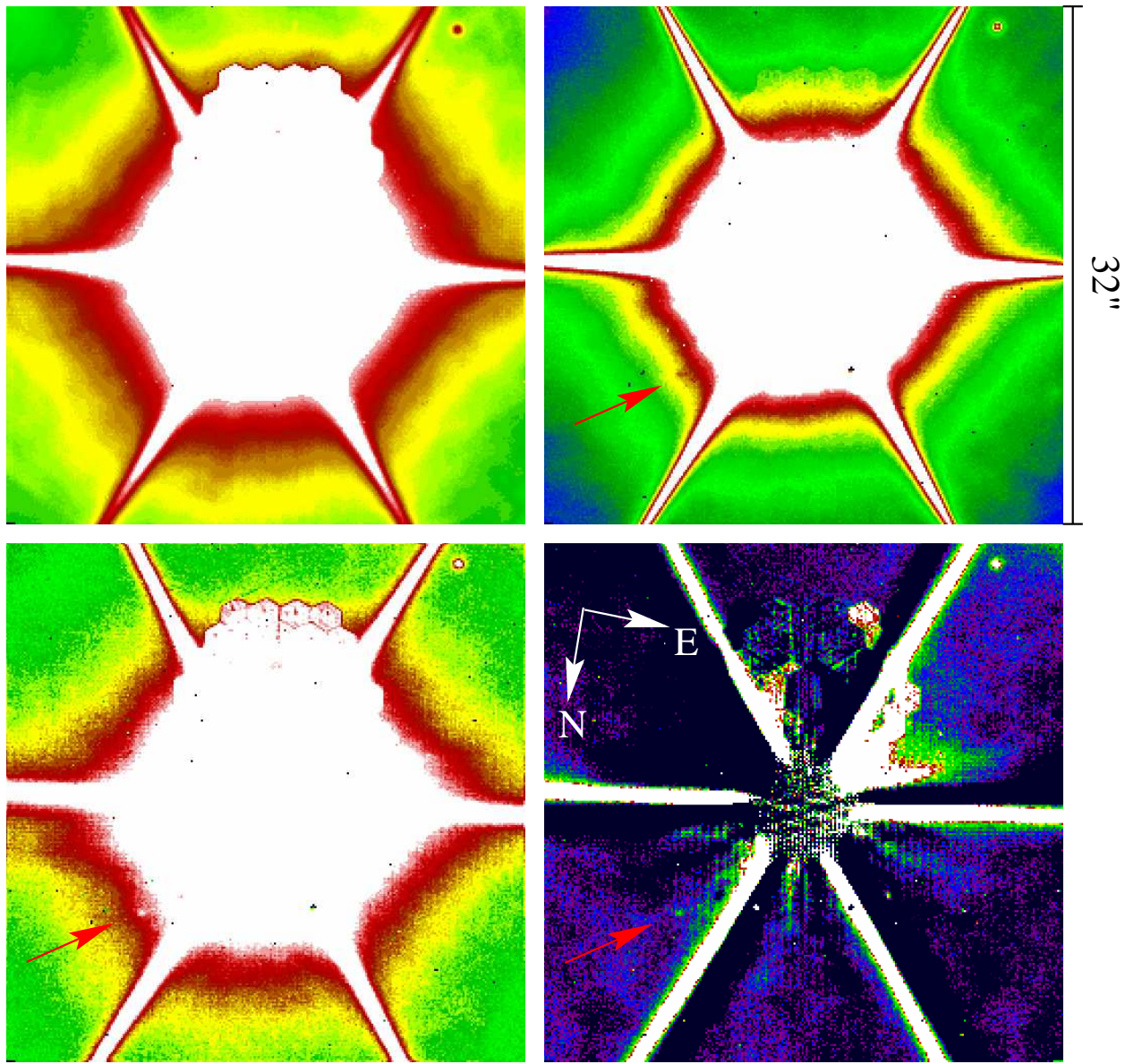}
\caption{Keck NIRC images of 64~Psc.  Left-to-Right, Top-to-Bottom the
images are $J$-band (1.2 $\mu$m), $H$-band (1.6 $\mu$m), $K$-band (2.2
$\mu$m), and a $K$ minus $K$-reflected image (see the text for an
explanation of this last image).  The field of view is 32\arcsec\ on a
side with a pixel size of 0.15\arcsec.  The red arrow indicates the
peculiar object, which we call 64~Psc~E, whose spectrum is shown in
Fig.~\ref{fig:spectrum}.  64~Psc~E's $J$ - $K$ color is $>$ 2.5$^{m}$.
\label{fig:64p_NIRCimages}}
\end{figure}

Because our 64 Psc orbital studies do not rule out the presence of a
companion as N79 suggest, we imaged the 64 Psc system with the
Near-Infrared Camera (NIRC) on the Keck-I telescope at Mauna Kea on
20 October 1998, and again on 28 November 1998.  A panel of
near-infrared images of the 64 Psc system is shown in
Figure~\ref{fig:64p_NIRCimages}.  

These images were obtained through the $J$, $H$ and $K$ filters with
total integration times of 60 s -- the data from six 10 s exposures
being coadded into a memory buffer before being recorded to disk.  In
the $J$ band we took 10 such images while moving the telescope
pointing by 10 pixels between each exposure to limit the effects of
bad pixels on the array.  These 10 images were then shifted and added.
The same procedure was used at $H$ and $K$ band, but only 7 images
were taken in $H$ and 1 in $K$.  These were sufficient to detect
objects down to $J$ = 21.0, $H$ = 20.2 and $K$ = 19.0, where the total
exposure times are 600 s, 420 s and 60 s for $J$, $H$ and $K$
respectively.  Clearly the presence of the bright 64~Psc system in the
images makes detecting faint objects difficult.  By reflecting the
images along the central vertical axis (where the star was positioned)
and subtracting this reflected image from the original, a substantial
portion of the star light is removed.  (See \cite{Matthews96} for a
detailed description of this technique.)  This subtracted image is
used for photometry of the remaining objects.  We used the standard
SJ9101 (HST P525-E, \cite{Persson98}) to calibrate the instrumental
magnitudes and determine the flux density of the point sources we
found in these images.

There are two apparent point sources other than 64~Psc itself in our
images, which following the WDS conventions we will refer to as 64 Psc
D and E (momentarily leaving aside the issue of whether these objects
are actually assocated with the 64~Psc system).  To determine the
offsets of the D and E objects with respect to 64~Psc we had to
determine the position of 64~Psc on the detector, which was completely
saturated within a radius of 23 pixels of the star.  (The pixel scale
is 0.15''.)  The star's position was determined by fitting lines
through the unsaturated portions of the 6 diffraction spikes yielding
a position with a conservative accuracy of better than 0.05'' (a third
of a pixel).  The first point source, herein 64 Psc D, is in the upper
right of the images, separated from 64 Psc by 22.7'' at a position
angle of 250$^\circ$, measured N through E.  This point source has the
following magnitudes: $J$ = 17.0 $\pm$ 0.1, $H$ = 16.8 $\pm$ 0.1 and
$K$ = 16.6 $\pm$ 0.1.  This photometry is consistent with that of a
background G7 V star at a distance of roughly 4~kpc.  It is therefore
apparently excluded as a possible companion to 64~Psc.

The other point source, herein 64 Psc~E, is much dimmer than the D
companion, and we have highlighted it in the figures with a red arrow
to indicate its position.  64~Psc~E is separated from 64 Psc by 12.1''
at a PA of 60$^\circ$, and is extremely red; it is undetected in the
$J$ image, but visible in $H$ and clearly detected at $K$ bands.
64~Psc~E shows no obvious signs of extended structure in either the
$H$ or $K$- band images.  Photometric measurements of 64~Psc~E yield
the following: $J$ $>$ 21.0, $H$ = 19.3 $\pm$ 0.2, $K$ = 18.5 $\pm$
0.15.  If this candidate is physically associated with 64~Psc it would
have an $M_K$ = 16.7 making it 2.4 magnitudes fainter than Gliese
229B, the coolest known brown dwarf and the least luminous condensed
object directly detected outside the solar system.

Because of 64~Psc~Es intriguing red color, and the fact that the
separation between it and 64 Psc is so near the calculated separation
if the N79 conjecture were correct, we obtained a near-infrared
spectrum with NIRC on the night of 28 November 1998.  

To obtain the spectrum we used NIRC's 120 line mm$^{-1}$ grism in
conjunction with the 4 pixel slit (chosen to match the seeing) and
blocking filters which permit only the $H$ and $K$ bands to pass
through to the detector.  The object was positioned on the slit and 14
exposures of 60~s integration (1 coadd) were taken, while moving the
object along the slit by 10 arcseconds between exposures.  This
permitted us to extract spectra from every exposure.  We then observed
the standard G8III star SAO 92183 to provide a spectral flatfield from
the same airmass.  SAO 92183 was smeared across the entire length of
the slit to produce a ``flatfield'' over the whole detector.  A sky
frame was taken identically to be used for subtraction from the
reference (SAO 92183) spectral image.  The data were reduced by
summing all the spectral images and dividing by the reference spectral
image.  Then the spectrum of 64~Psc~E and that of one of the diffraction
spikes of 64 Psc were extracted separately.  The two were separated by
5'' in the spectral images.  This means that it is unlikely any
contamination from 64 Psc is present in the spectrum of 64~Psc~E.  The
spectra were extracted by summing the 7 rows over which the signal
from each source was visible.  The extracted spectra were then
wavelength calibrated using the well-determined 50\%\ transmission
points of the $HK$ blocking filter.  The flux calibration involved
first the removal of the Rayleigh-Jeans falloff of the reference
G-star used to flat field the data and a matching of the $K$-band
integrated flux to that measured by the photometry.

\begin{figure}[p]
\epsscale{0.8}
\plotone{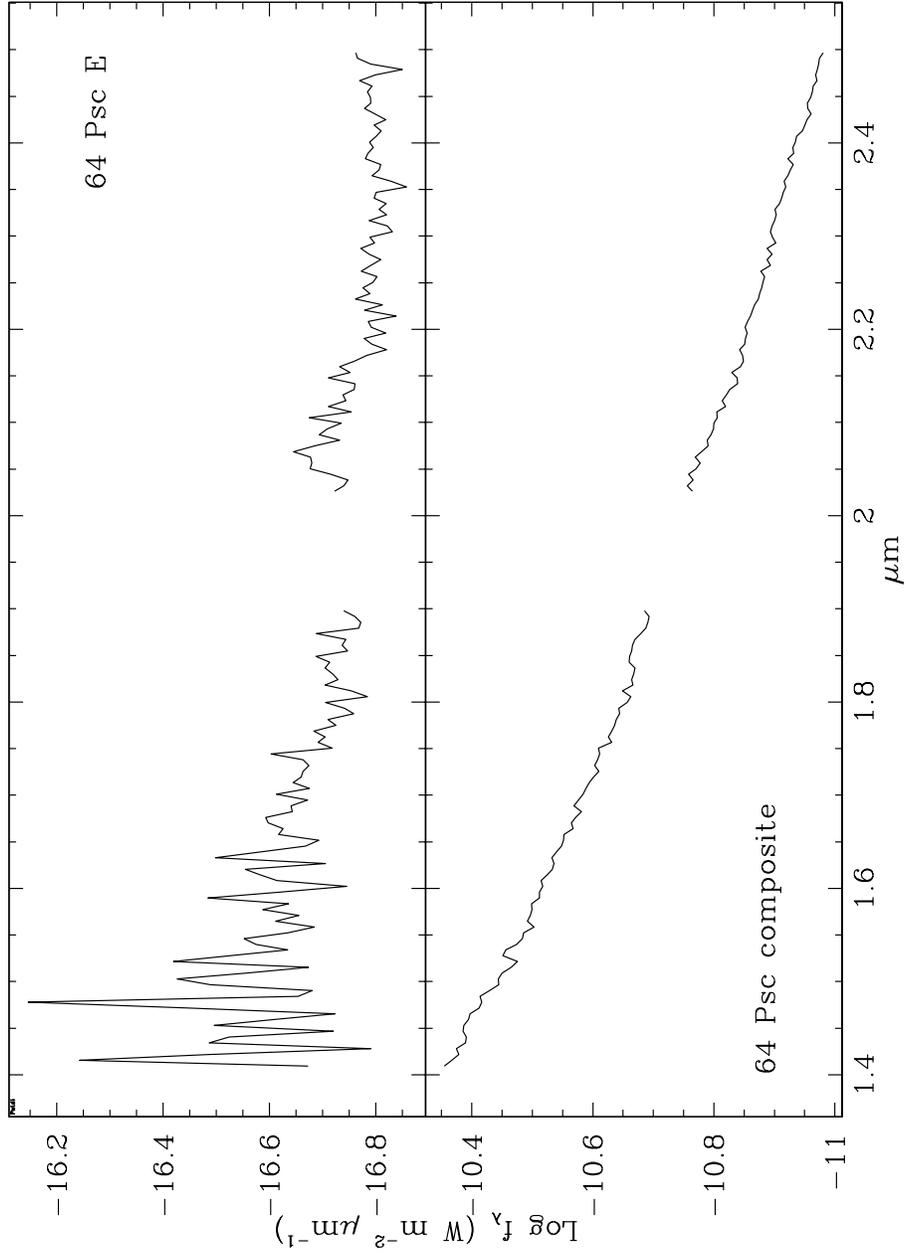}
\caption{The near-infrared spectrum of the candidate companion of 64
Psc, 64~Psc~E, is shown in the top panel.  The bottom panel is the
composite spectrum of the two F stars that comprise 64 Psc.  That
spectrum was taken from the diffraction spike which passed through the
slit of the spectrograph. The gap in both spectra between 1.9 and 2.0
$\mu$m is where the atmospheric absorption prevented measurement.  We
estimate the SNR in the 64~Psc~E spectrum is approximately 2 in the
blue edge of the $H$-band and 35 in the red end of the $K$-band.
\label{fig:spectrum}}
\end{figure}

The spectra from 1.4 to 2.5 $\mu$m of both the candidate companion
64~Psc~E and 64 Psc itself are shown in Figure~\ref{fig:spectrum}.  The
candidate spectrum is extremely red, confirming the photometric
measurements, and it is substanitally different from the composite
spectrum of the two late F-star components of 64 Psc.  This
demonstrates that little or no contamination from 64~Psc itself is
present in the spectrum of 64~Psc~E.  While the SNR is low, there are no
apparent molecular features in the spectrum.  In particular the large
methane features seen in the spectrum of the cool brown dwarf GL 229B
(see \cite{Oppenheimer95,Geballe98,Oppenheimer98}) are notably absent.

Based on the colors and proximity of the first (64~Psc~D) candidate,
and the implied large absolute magnitude of the second (64~Psc~E)
candidate, we think it is unlikely that either object is physically
associated with the 64 Psc system.  However, 64~Psc~E is relatively
unusual in its extreme red colors.  Based on initial statistics from
the 2MASS survey, L-dwarfs typically have $J$ - $K$ colors less than
2.1, so that possibility would seem to be unlikely
(\cite{Kirkpatrick99}).  Other possible explanations for the extreme
red nature of 64~Psc~E is the possibility of a very red QSO
(\cite{Beichman98}), or a red, compact background galaxy
(\cite{Becklin95,Graham96,Stanford97}).  It is clearly interesting to
determine whether the 64~Psc~E object shows indications of common
proper motion with 64~Psc itself, and a higher SNR spectrum of the
object would be needed to unequivocally establish the nature of
64~Psc~E.

\section{Summary and Discussion}
By virtue of our interferometric resolution and the high precision of
the DM91a radial velocity data we are able to determine accurate
masses of the 64~Psc constituents, and an accurate system distance.
Our data favors the earlier spectroscopic period of N79 over the more
recent DM91b determination.  The approximate determination of relative
K-brightness ratio of the two nearly-equal brightness 64~Psc
constituents (best-fit K-magnitude difference of 0.141 $\pm$ 0.076)
prevents us from establishing tight constraints on the absolute
magnitudes of the components.  This is regrettable, but is an inherent
limitation of our amplitude-only reconstruction technique for nearly
equal-brightness systems (\cite{Hummel98}).

The N79 suggestion of an additional body in the system as an
explanation for purported orbital parameter variation is intriguing,
but our orbital results do not seem to favor this inference.  However,
these considerations suggest high angular resolution, high dynamic
range imaging of the system to attempt a detection of a companion, and
we have imaged the 64 Psc system with NIRC at Keck-I.  Neither of the
two potential companion candidates apparent in our near-infrared
images seem likely to be physically associated with the 64 Psc system.
However, the second of these candidates (herein 64~Psc~E) would appear
to be fairly unusual based on its very red colors and spectrum in the
near-infrared, and follow-up proper-motion and spectroscopic
observations would be warranted.

\acknowledgements The work described in this paper was performed at
the Jet Propulsion Laboratory and the Infrared Processing and Analysis
Center, California Institute of Technology under contract with the
National Aeronautics and Space Administration.  Interferometer data
was obtained at the Palomar Observatory using the NASA Palomar Testbed
Interferometer, supported by NASA contracts to the Jet Propulsion
Laboratory.  Imaging and spectroscopic data was obtained at the
W.M.~Keck Observatory at Mauna Kea, Hawaii, operated by the California
Association for Research in Astronomy.  We thank K.Y.~Matthews (CIT)
for assistance with near-infrared photometry on 64 Psc.  A.F.B.~in
particular thanks C.A.~Hummel (USNO) for his suggestion concerning
integrated fitting of interferometric visibilities and radial
velocities.

We wish to thank the anonymous referee for his many positive
contributions to the accuracy and quality of this manuscript.

This research has made use of the Simbad database, operated at CDS,
Strasbourg, France.

\end{document}